# Computational Pathology for Accurate Prediction of Breast Cancer Recurrence: Development and Validation of a Deep Learning-based Tool

Ziyu Su[a*], Yongxin Guo[a], Robert Wesolowski[b], Gary Tozbikian[c], Nathaniel S. O'Connell[d], M. Khalid Khan Niazi[a], Metin N. Gurcan[a]

a. Center for Artificial Intelligence Research, Wake Forest University School of Medicine, Winston-Salem, NC, USA;

b. Division of Medical Oncology, Comprehensive Cancer Center, The Ohio State University College of Medicine, Columbus, OH, USA;

c. Department of Pathology, The Ohio State University, Columbus, OH, USA;

d. Department of Biostatistics and Data Science, Wake Forest University School of Medicine, Winston-Salem, NC, USA

**\*Corresponding author:**

E-mail: zsu@wakehealth.edu

## Abstract

Accurate recurrence risk stratification is crucial for optimizing treatment plans for breast cancer patients. Current prognostic tools like Oncotype DX (ODX) offer valuable genomic insights for HR+/HER2- patients but are limited by cost and accessibility, particularly in underserved populations. In this study, we present Deep-BCR-Auto, a deep learning-based computational pathology approach that predicts breast cancer recurrence risk from routine H&E-stained whole slide images (WSIs). Our methodology was validated on two independent cohorts: the TCGA-BRCA dataset and an in-house dataset from The Ohio State University (OSU). Deep-BCR-Auto demonstrated robust performance in stratifying patients into low- and high-recurrence risk categories. On the TCGA-BRCA dataset, the model achieved an area under the receiver operating characteristic curve (AUROC) of 0.827, significantly outperforming existing weakly supervised models (p=0.041). In the independent OSU dataset, Deep-BCR-Auto maintained strong generalizability, achieving an AUROC of 0.832, along with 82.0% accuracy, 85.0% specificity, and 67.7% sensitivity. These findings highlight the potential of computational pathology as a cost-effective alternative for recurrence risk assessment, broadening access to personalized treatment strategies. This study underscores the clinical utility of integrating deep learning-based computational pathology into routine pathological assessment for breast cancer prognosis across diverse clinical settings.

## 1. Introduction

Breast cancer is the most prevalent cancer and the second biggest reason for cancer-related death in women in the United States [1]. The effective treatment options and prognosis for breast cancer patients are highly dependent on the patient's molecular subtype of breast cancer as determined by estrogen, progesterone, and human epidermal growth factor 2 (HER2) receptor expression. Among all different subtypes, hormone receptor-positive (HR+) and epidermal growth factor receptor-negative (HER2-) breast cancer represents the most common entity, accounting for approximately 65% of all cases [2, 3]. This subtype of breast cancer is highly dependent on estrogen signaling, and most such patients can be safely treated with endocrine therapy aimed at blocking this signaling. However, chemotherapy effectively reduces the recurrence risk in a small subset of patients with operable disease. At the same time, it does not significantly benefit the majority of this group. Given the significant short- and long-term toxicities of chemotherapy that can badly affect patients' quality of life, it is critical to know which patients will derive significant benefits from cytotoxic chemotherapy to justify these risks.

The Oncotype DX (ODX) test is widely used as a prognostic and predictive tool and a prognosis indicator for breast cancer recurrence risk in patients with operable HR+/HER2- breast cancer. It is a gene-assay based on 21 critical gene expression levels obtained through reverse transcriptase-polymerase chain reaction (RT-PCR) technology [4]. The assay assigns a risk score ranging from 0 to 100, with a higher score indicating a higher risk of breast cancer recurrence. Recurrence risk is also an indicator suggesting the benefits of chemotherapy. Recent research reports that a risk score of 25 or less is associated with no benefit from receiving adjuvant chemotherapy for women with node-negative, HR+/Her2- above who are older than 50 years old or post-menopausal women with the presence of metastases to 1-3 axillary lymph nodes [5, 6]. On the other hand, for women under 50 years old with early-stage, node-negative, HR+/HER2- breast cancer, and ODX recurrence score between 16-25, there are modest chemotherapy benefits and the decision whether to use it is based on close assessment of other risk factors and thorough discussion between healthcare provider and the patient. Therefore, a common practice is stratifying patients into low or high-risk based on a cut-off value 25.

Although effective, ODX has limited usage for the vast majority of breast cancer patients due to its high cost (~4000 US dollars per test). According to the World Bank statistics [7], 70 countries with low-income or lower-middle-income economies have average annual per capita incomes of less than $1,146 and $4,515, respectively. It would be difficult, if not impossible, for women in these countries and several other countries to access expensive tests. Several studies have shown that race/ethnic minorities in the U.S., such as African Americans and Hispanic Americans, have fewer chances to receive ODX compared to Caucasian Americans [8-10]. Furthermore, inner-city and municipal hospitals have lower ODX utilization than private hospitals [11]. Therefore, there are significant racial and income disparities in the utilization of ODX, highlighting the need for more accessible and affordable prognostic tools to ensure equitable healthcare for all breast cancer patients [12].

Recently, several studies have explored machine learning methods to predict ODX outcomes using readily accessible medical data modalities [12-15]. Among these approaches, computational pathology has demonstrated significant potential as a cost-effective alternative to gene assays, thanks to the affordability of histopathology slides and advancements in deep learning technology [13, 14, 16]. Computational pathology involves the computational analysis of digitized histopathology slides, namely the whole slide images (WSIs) [17]. With the progress in deep learning, it is now possible to extract quantitative features from WSIs, enabling the prediction of clinically significant outcomes.

Previous studies have demonstrated the potential of predicting ODX outcomes from H&E-stained WSIs using deep learning techniques [18-21]. These approaches leverage deep learning models to extract hand-crafted nuclei features from WSIs. However, developing these methods requires massive annotation efforts from pathologists, which is costly. In addition, their multi-stage feature extraction methods increase the implementation difficulties and variabilities. With the advancements in the deep learning domain in the past five years, weakly supervised learning models have gradually become the mainstream of deep learning models for computational analysis. According to the weakly supervised learning paradigm, deep learning models are not used to extract hand-crafted features. Instead, they treat a high-resolution WSI as a single input and automatically learn clinically relevant features from WSIs. Multiple successful applications of weakly supervised learning have emerged in various tasks, such as tumor metastasis classification [22, 23], subtyping [24], survival prediction [25], etc. However, the performance of existing methods on ODX prediction has remained moderate [14]. Our latest study showed promising results, achieving an AUC higher than 0.8 [13]. Nevertheless, this assessment was based on a relatively small dataset comprising 150 WSIs, indicating the need for validation on a larger cohort to confirm these findings and improve the model's robustness.

In this study, we introduce a fully automatic pipeline called Deep-BCR-Auto, which utilizes deep learning models to predict ODX results based on H&E-stained WSIs. The pipeline comprises two main components: tumor bulk segmentation and weakly supervised ODX prediction, eliminating the need for pathologists' annotations. Our dataset is sourced from the TCGA Breast Cancer (BRCA) database, encompassing a total of 1,006 patients, including 516 with HR+/HER2- status. Research-based ODX recurrence risks are calculated from the provided patients' mRNA expression data [16]. Our experimental results demonstrate that Deep-BCR-Auto accurately predicts ODX-based breast cancer recurrence risk from WSIs. The model's performance remains consistently robust across different racial and age groups, demonstrating its potential for broad clinical application. When evaluated independently on an in-house dataset with actual ODX results, Deep-BCR-Auto maintained strong generalizability. Our approach not only offers a cost-effective and accessible alternative to traditional gene assays but also addresses disparities in healthcare access by providing reliable prognostic information to a diverse patient population.

## 2. Methods
### 2.1. Dataset

Our study utilizes two datasets of H&E-stained breast cancer resections: the publicly available TCGA Breast Cancer (TCGA-BRCA) dataset and an in-house dataset referred to as the OSU dataset. Details of each dataset are outlined below.

The TCGA-BRCA dataset consists of 1133 WSIs, which are employed for training and testing the deep learning models. Although the original dataset does not include ODX results, we utilized the research-based ODX results calculated by Howard et al. [16], which are derived from the TCGA provided patient-level mRNA expression data. In a nutshell, they normalized the sequencing data and calculated the score using the ODX formulas. Subsequently, they chose a threshold differentiating high/low-risk patients based on the 15$^{th}$ percentile results of all HR+/HER2- patients, following the statistics in the National Cancer Database [16, 26].

We performed data cleaning to remove WSIs from TCGA-BRCA dataset with missing ODX or receptor status information, as well as those that failed to undergo processing in our slide processing pipeline. As a result, the finalized TCGA-BRCA dataset comprises 1065 H&E-stained WSIs sourced from 1006 patients (i.e. some patients have multiple WSIs). Among these patients, 516 patients are HR+/HER2-, with 443 classified as low risk and 73 classified as high risk. Detailed demographics are summarized in Table 1. In our study, we primarily use HR+/HER2- patient data for model training and testing. For patients who are not HR+/HER2-, we include their WSIs in our training set to enhance the training of the deep learning models (details can be found in Section 2.4).

**Table 1. Demographics of HR+/HER2- patients in the TCGA-BRCA dataset.** In this table, n indicates the total number, and SD indicates standard deviation.

|  |  | Low-risk | High-risk |
|---|---|---|---|
| Total, n |  | 443 | 73 |
| Age, mean (SD) |  | 64.9 (9.8) | 64.6 (11.0) |
| Sex, n (%) | Female | 440 (99.3) | 73 (100) |
|  | Male | 3 (0.7) | 0 (0) |
| Race, n (%) | Asian | 17 (3.8) | 4 (5.5) |
|  | Black | 34 (7.7) | 11 (15.1) |
|  | White | 346 (78.1) | 43 (58.9) |
|  | Missing | 46 (10.4) | 15 (20.5) |
| Histologic type, n (%) | Lobular | 86 (19.4) | 5 (6.8) |
|  | Ductal | 249 (56.2) | 55 (75.4) |
|  | Lobular & Ductal | 64 (14.4) | 7 (9.6) |
|  | Other | 44 (10.0) | 6 (8.2) |
| Grade, n (%) | 1 | 150 (33.9) | 8 (11.0) |
|  | 2 | 211 (47.6) | 21 (28.8) |
|  | 3 | 82 (18.5) | 44 (60.2) |

The OSU dataset is an in-house dataset collected from The Ohio State University Wexner Medical Center (OSUWMC). This study is IRB-approved by the Ohio State University Cancer Institutional Review Board, with a Waiver of the Consent Process and Full Waiver of HIPAA Research Authorization. It consists of 465 H&E-stained WSIs of tumor resections from HR+/HER2- breast cancer patients. In total, the OSU dataset consists of 384 WSIs from ODX≤25 (low-risk) cases and 81 WSIs from ODX>25 (high-risk) cases. In our study, we used the OSU dataset as an independent testing set to evaluate the accuracy and generalizability of our deep learning model trained on the TCGA-BRCA dataset.

### 2.2. Automatic ODX prediction pipeline

In this study, we developed a fully automatic WSI-based ODX prediction pipeline, DeepBCR-Auto. The proposed approach for ODX prediction includes two main steps: 1) tumor bulk segmentation and 2) ODX prediction on tumor bulks (see Figure 1).

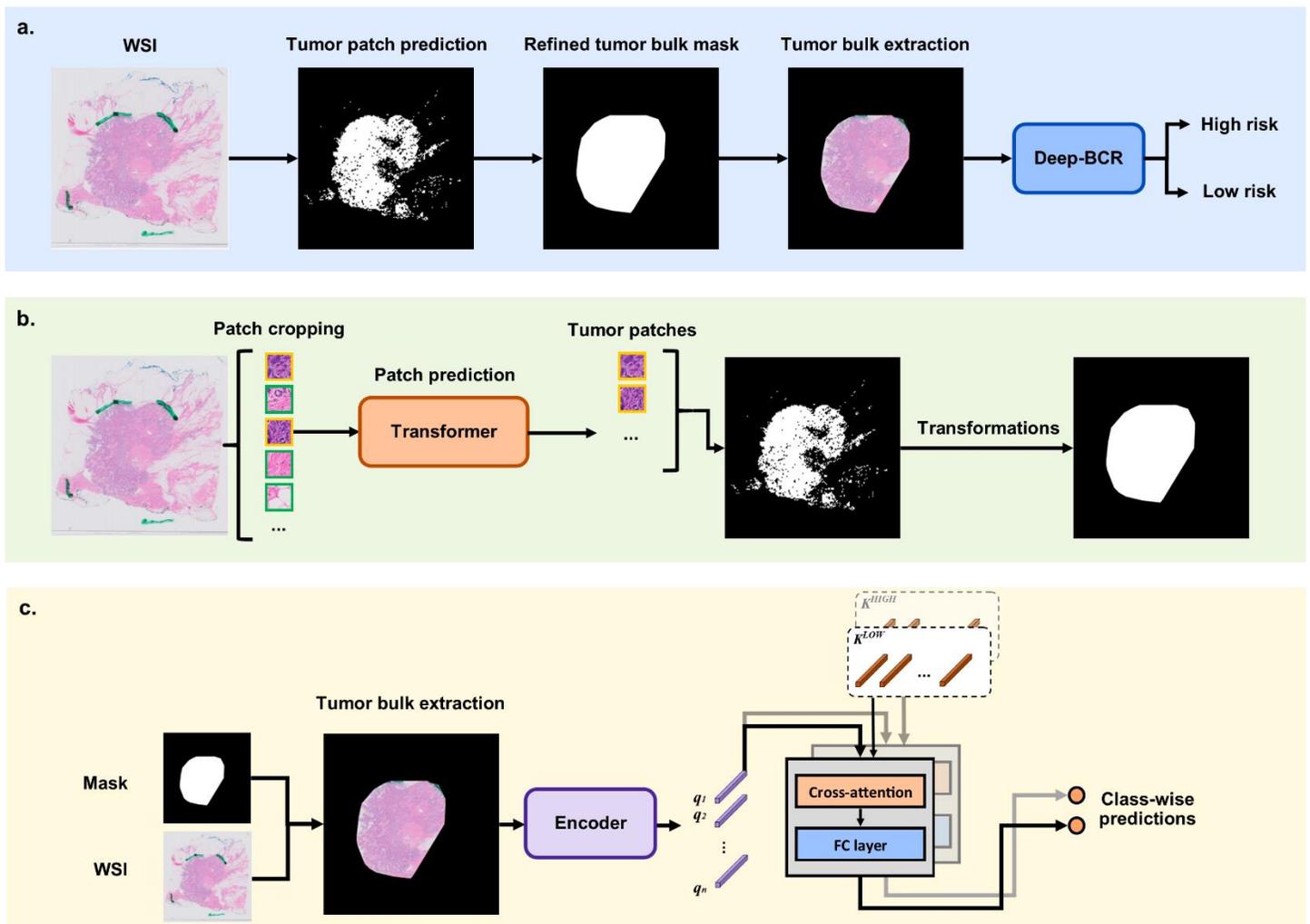

**Figure 1. Schematics of Deep-BCR-Auto. a.** Overview of the Deep-BCR-Auto pipeline, including tumor patch prediction, tumor bulk mask refinement, tumor bulk extraction, and ODX prediction. **b.** Details of the tumor bulk segmentation procedure. A Transformer-based model is trained as a patch classifier to distinguish tumor patches from input WSIs. The predictions are mapped to a binary mask at the slide level and then refined to extract the main tumor bulks using morphological transformations. **c.** Deep-BCR architecture. Deep-BCR is a weakly supervised learning model designed to stratify patients into low or high recurrence risk based on the tumor bulk regions of the WSIs.

As a preprocessing step for our pipeline, we first performed foreground segmentation to extract tissue regions from each WSI using the color thresholding method [27]. Following the common practice in slide-level histopathology analysis [17, 25, 28], we then cropped the WSIs into image patches at size 896×896 pixels under 40× magnification.

To remove the tumor-irrelevant regions from the WSIs, we applied a slide-level tumor bulk segmentation model on each breast cancer WSI. Tumor bulks are the main tumor regions on the resections that mainly comprise tumor cells along with some intratumoral and peritumoral stroma. For this step, we developed a patch classifier neural network composed of a pre-trained Transformer model [29], a dropout layer, and a randomly initialized linear classification layer.

To train this model, we utilized our internal breast cancer resection dataset [14]. An experienced pathologist annotated the tumor bulk for 50 H&E-stained WSIs. We then randomly cropped 36077 tumor patches and 36077 normal patches to train our patch classifier. During the training, we made the last transformer block and the linear classification layer tunable, and the rest of the model was frozen. As a result, this model achieved an 89.2% tumor/normal patch classification accuracy on a small leave-out testing set (details in Supplemental Methods 1).

Following the development of the patch classifier, we applied it to the TCGA-BRCA and OSU WSIs to detect the tumor patches. Before applying the classifier, we performed color normalization on all the input patches [30] to match the color style across different datasets. We mapped the prediction results on a binary mask for each WSI. To refine the tumor bulk mask, we performed a series of morphological transformations, including binary closing, small object removal, and convex hull. Subsequently, we only preserve the two largest segmented tumor areas for each WSI.

Once the tumor bulk was segmented, we kept the patches from the tumor bulk only. Then, we encoded the patches into feature vectors using a histopathology foundation model [29]. Finally, based on the encoded WSIs, we trained Deep-BCR, a weakly supervised learning model proposed in our previous study [13], to predict high/low-risk for breast cancer recurrence. The details of this weakly supervised model will be introduced in the following section.

### 2.3. Weakly supervised learning model

Weakly supervised learning, also known as multiple instance learning (MIL), is a machine learning paradigm that aims to classify a bag of instances instead of individual instance [17]. During the training, the model can only see the bag-level labels without seeing the label for each instance. Thus, the model needs to comprehensively analyze and aggregate the instance-level features to make a bag-level prediction. The weakly supervised learning model is widely applied in WSI analysis due to its label-efficient property [17]. Given the giga-pixel level resolution of WSIs, it is challenging for pathologists to extensively annotate each region of a WSI, which results in limitations of carefully annotated WSI datasets. For some tasks, such as our recurrence risk prediction problem, it is impossible for the pathologists to indicate which regions are responsible for high ODX scores. With the help of weakly supervised learning, we can formulate a WSI as a bag and encoded patches as instances, so that we only need to collect the slide-level or patient-level outcomes to train our model. The model will intelligently aggregate the patch features to predict the slide-level outcomes.

In our pilot study [13], we proposed Deep-BCR, a weakly supervised learning model for ODX recurrence risk prediction. The essential idea of Deep-BCR is to learn attention scores for each input patch of a WSI and then aggregate all the encoded patches into a slide-level representation using the learned attention scores. The slide-level representation will be used to predict the recurrence risk. Unlike other models, Deep-BCR employs a cross-attention neural network to correlate each input patch with some representative patches from both low and high-risk categories. The correlations are then used to learn the attention weights for the patches via a neural network. The architecture of Deep-BCR is depicted in Figure 1c and more details can be found in [13]. The implementation details for training Deep-BCR are described in Supplemental Methods 2.

### 2.4. Experimental design

Our task is formulated as an image classification problem to determine low or high-risk categories. To develop our model, we performed a stratified three-fold cross-validation on the HR+/HER2- cohort of the TCGA-BRCA dataset. The dataset was split at the patient level to ensure no overlap of patient data between the training and testing sets. Due to the categorical imbalance of the dataset, we augmented the training set for each fold with all the non-HR+/HER2- data to create a relatively balanced training set.

To thoroughly test the accuracy and generalizability of Deep-BCR-Auto, we used the OSU dataset as an independent testing set for model trained on the TCGA-BRCA dataset. For independent testing, we picked the model performed the best under three-fold cross-validation on the TCGA-BRCA dataset.

To evaluate our model, we report AUROC (Area Under the Receiver Operating Characteristic) and AUPRC (Area Under the Precision-Recall Curve). The 95% confidence intervals (CI) were calculated using the 1000-time bootstrapping method. For comparisons, we selected CLAM [27] and CLAM-CTrans as the comparison methods in our study, as they demonstrated excellent performance in our previous ODX prediction study [13]. CLAM is a widely used MIL model that has been applied across various WSI analysis tasks [31-33], while CLAM-CTrans is the improved version of CLAM, incorporating a histopathology-specific patch encoder [29]. P-values for AUROC were calculated to compare Deep-BCR-Auto against the comparison models using Delong's test. Additionally, we adjusted our prediction threshold to achieve 70%, 80%, and 90% sensitivities to test the rule-out performance of our model under high sensitivities. Performance was assessed using confusion matrices, specificity, negative predictive value (NPV), and positive predictive value (PPV).

## 3. Results

In this section, we first report the cross-validation performance of the Deep-BCR-Auto compared with other MIL models on the TCGA-BRCA dataset (see Section 3.1). Then, we selected the best-performing model on the TCGA-BRCA dataset and tested it on our in-house OSU dataset (see Section 3.2).

### 3.1 Cross-validation results on TCGA-BRCA dataset

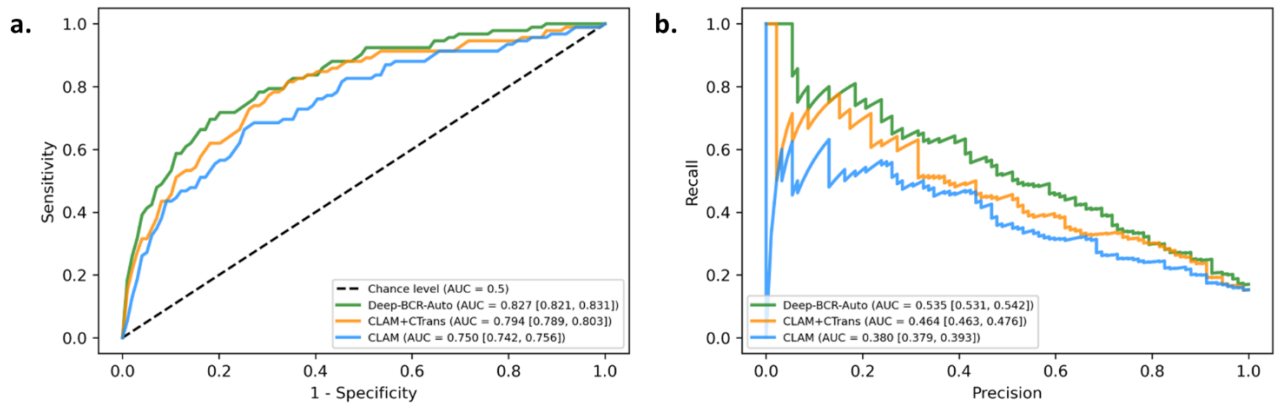

**Figure 2. Prediction curves for recurrence risk on TCGA-BRCA dataset. a.** Receiver operating characteristic curves for Deep-BCR and comparison weakly supervised models. **b.** Precision-Recall curves for Deep-BCR and comparison weakly supervised models. AUC are reported in the legends accompanied with 95% CI in brackets []. The 95% CIs are calculated using the 1000-time bootstrapping method across the three-fold cross-validation.

**Table 2. Prediction performance for recurrence risk on TCGA-BRCA dataset.** The 95% CI are reported in brackets [] based on 1000-time bootstrapping across the three-fold cross-validation. p-values are calculated to compare Deep-BCR-Auto against comparison models using Delong's test.

|  | AUROC [95% CI] | AUPRC [95% CI] | p-value (Compared to Deep-BCR-Auto) |
| --- | --- | --- | --- |
| **Deep-BCR-Auto** | **0.827 [0.821, 0.831]** | **0.535 [0.531, 0.542]** | - |
| CLAM+CTrans | 0.794 [0.789, 0.803] | 0.464 [0.463, 0.476] | 0.366 |
| CLAM | 0.750 [0.742, 0.756] | 0.380 [0.379, 0.393] | 0.041 |

In Figure 2 and Table 2, we report the performance of Deep-BCR-Auto in predicting ODX recurrence risk using H&E-stained WSIs from the TCGA-BRCA dataset. We compared Deep-BCR-Auto with CLAM [27] and CLAM-CTrans. As a result, the Deep-BCR-Auto achieved the best performance (AUROC 0.827, 95% CI 0.821-0.831) that outperforms CLAM-CTrans (AUROC 0.794, 95% CI 0.789-0.803) and CLAM (AUROC 0.750, 95% CI 0.742-0.756). Notably, Deep-BCR-Auto was significantly better than the original CLAM based on the Delong's test (p=0.041). This result indicates superior accuracy of the proposed Deep-BCR-Auto model.

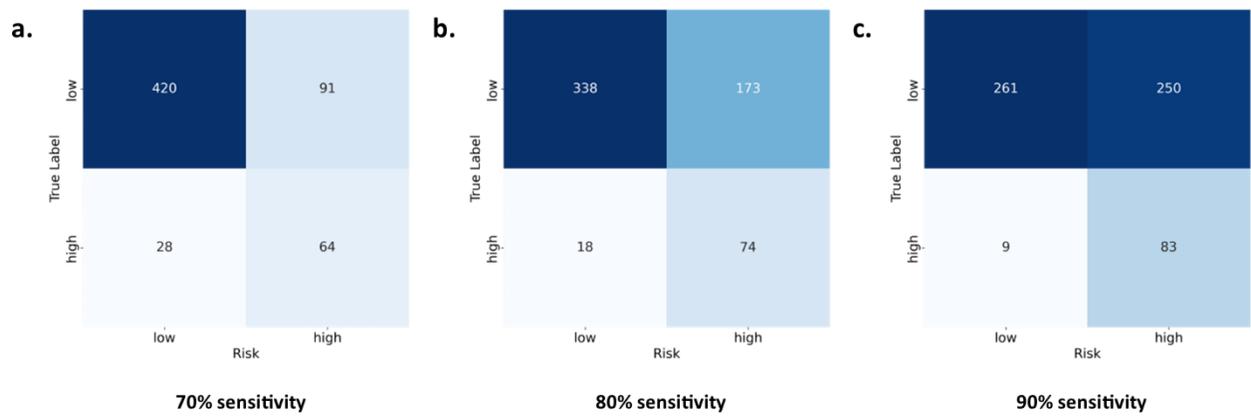

**Figure 3. Confusion matrices of Deep-BCR-Auto with different thresholds adjusted for high sensitivities on TCGA-BRCA dataset.**

**Table 3. Prediction performance of Deep-BCR-Auto with different thresholds adjusted for high sensitivities on TCGA-BRCA dataset.**

| Sensitivity | Specificity | PPV | NPV |
|---|---|---|---|
| 70.0% | 82.2% | 41.3% | 93.6% |
| 80.0% | 66.1% | 30.0% | 94.9% |
| 90.0% | 51.1% | 24.9% | 96.7% |

In clinical practice, a high-sensitivity model is crucial to identifying the majority of high-risk patients. Therefore, we adjusted Deep-BCR-Auto's prediction threshold to achieve 70%, 80%, and 90% sensitivities. The resulting performance is concluded in Figure 3 and Table 3. We found that our model maintains reasonable specificities while achieving high sensitivities. This indicates that our model can effectively rule out low-risk patients, potentially reducing the use of unnecessary genetic tests.

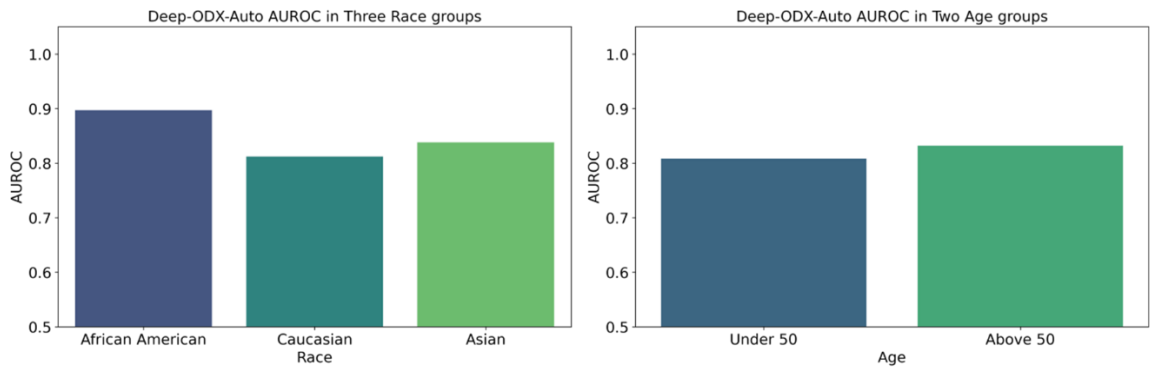

**Figure 4. AUROC reported in different racial and age groups on TCGA-BRCA dataset.**

In Figure 4, we also reported Deep-BCR-Auto's AUROC in different races and age groups. Our model achieved AUROCs of 0.898, 0.813 (p=0.123), and 0.838 (p=0.58) in African American, Caucasian, and Asian race groups, respectively (see Figure 4a). This result indicates that our model can yield comparable prediction performance among different race groups, which reveals potential for broader applicability in diverse populations. This is crucial in clinical settings, where healthcare equity and the effectiveness of diagnostic tools across various demographic groups are paramount. In addition, our model also achieved similar performance between patients above and under 50 years old, with AUROCs of 0.808 and 0.832 (p=0.579), respectively. Overall, our model exhibits robustness across a wide demographic spectrum.

**3.2 Independent testing on the OSU dataset**

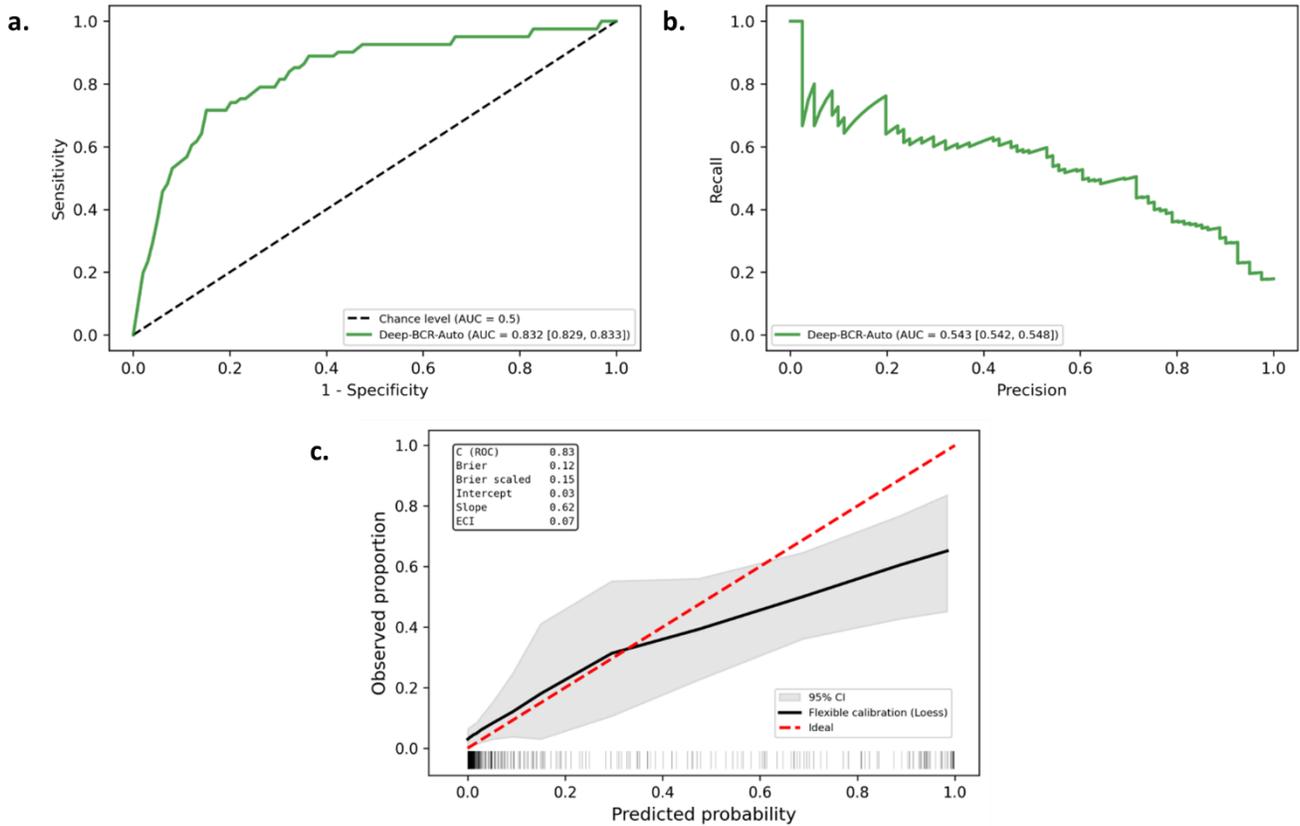

**Figure 5. Prediction curves of Deep-BCR-Auto on OSU dataset. a.** Receiver operating characteristic curves for Deep-BCR-Auto. **b.** Precision-Recall curves for Deep-BCR-Auto. The model is trained on TCGA-BRCA dataset and tested on the OSU dataset. AUC are reported in the legends accompanied with 95% CI in brackets []. The 95% CIs are calculated using the 1000-time bootstrapping method. **c.** Calibration plot illustrating Deep-BCR-Auto's performance in predicting breast cancer recurrence. The x-axis shows predicted probabilities, and the y-axis indicates observed high-risk rates. The solid black line is the LOESS-based calibration curve, with the red dashed line representing perfect calibration. The shaded area represents the 95% confidence interval. A rug plot at the bottom shows the distribution of patient predictions.

**Table 4. Prediction performance of Deep-BCR-Auto on OSU dataset.** The model was trained and tested on the TCGA-BRCA dataset in three-fold cross-validation. The best performing model was selected and tested on the OSU dataset. The classification threshold is tuned based on a validation set held out from the OSU dataset.

| Dataset | Accuracy | Specificity | Sensitivity | PPV | NPV |
|---|---|---|---|---|---|
| TCGA-BRCA testing set (the best performed fold) | 78.9% | 79.5% | 75.8% | 41.7% | 94.4% |
| OSU testing set (independent testing) | 82.0% | 85.0% | 67.7% | 48.4% | 92.6% |

For the Deep-BCR-Auto trained on the TCGA-BRCA dataset, we performed an independent testing experiment on the OSU dataset. Out of the three models we trained in the three-fold cross-validation, we selected the best performing model for the TCGA-BRCA dataset. Then, we applied the selected model to the WSIs from the OSU dataset. In Figure 5, we plot the receiver operating characteristic (ROC) and the precision-recall (PR) curves along with the AUCs of this experiment. We found that the proposed Deep-BCR-Auto achieved an AUROC of 0.832 (95% CI 0.829-0.833), which is comparable to its performance on the TCGA-BRCA dataset (AUROC 0.827, 95% CI 0.821-0.831).

The calibration performance of Deep-BCR-Auto, as shown in Figure 5c, highlights its clinical utility in predicting ODX recurrence risk. The calibration curve (solid black line) represents the relationship between predicted high-risk probabilities and observed proportions. In a perfectly calibrated model, predictions would fall directly on the ideal line (dashed red line), indicating that the predicted risk matches the true recurrence risk. For Deep-BCR-Auto, the calibration curve follows the ideal line closely at lower risk levels, indicating that the model accurately predicts low-risk patients. However, at higher predicted probabilities, the curve diverges, reflecting slight underestimation of risk in higher-risk patients. With a C (ROC) of 0.83, the model demonstrates strong discrimination between high- and low-risk patients, which is crucial for stratifying treatment decisions. However, the slope of 0.62 indicates slight underestimation in high-risk cases, a potential area for refinement to improve treatment allocation in patients who may require more aggressive intervention.

Since Deep-BCR-Auto was trained and tested on datasets from different hospitals, we further adjusted the classification threshold using a small validation set held out from the OSU dataset. Specifically, we split 15% of the WSIs for validation (denoted as OSU validation set) and 85% for testing (denoted as OSU testing set). We then selected the threshold that maximized the model's F1-score on the OSU validation set, balancing accuracy between the two categories. Testing was subsequently conducted on the OSU testing set using the optimized threshold, with results shown in Table 4. The proposed model achieved an overall accuracy of 82.0%, with 85.0% specificity and 67.7% sensitivity, comparable to its performance on the TCGA-BRCA dataset (accuracy 78.9%, specificity 78.5%, and sensitivity 75.8%). These findings highlight the excellent generalizability of Deep-BCR-Auto across datasets from different hospitals.

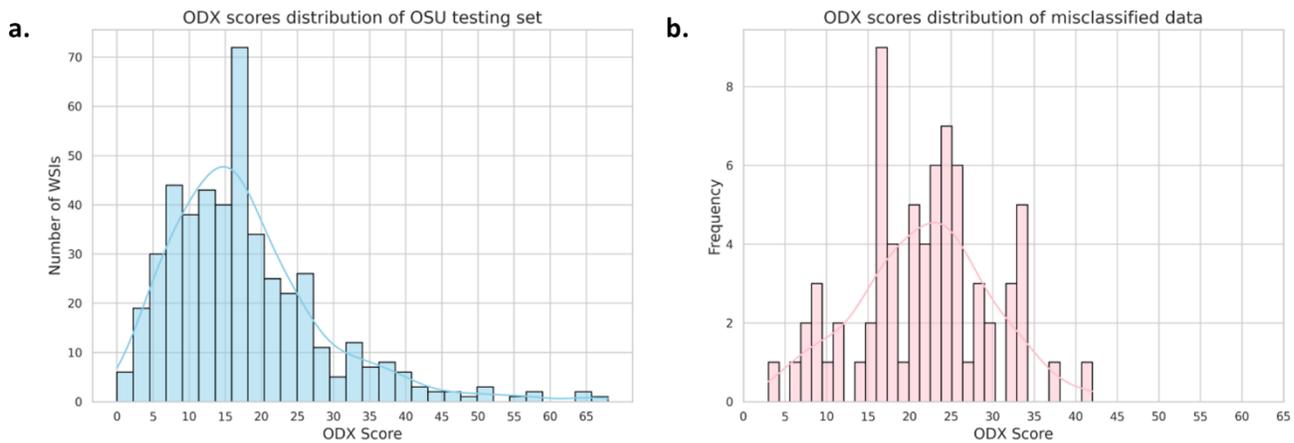

**Figure 6. ODX score analysis of misclassified samples from the OSU testing set. a.** The overall ODX score distribution of the OSU testing set. **b.** The ODX score distribution of misclassified samples from the OSU testing set. Curves in the plots indicate the smoothed distributions.

We further analyzed the Deep-BCR-Auto's misclassified samples from the OSU testing set. In Figure 6, we compare the ODX score distribution of misclassified samples to that of all samples. We found that misclassified samples primarily concentrated in the 20 to 30 score range, while samples with extremely low or high scores are less likely to be misclassified. As a reference, the original sample distribution is skewed towards lower ODX scores. These findings suggest that samples with intermediate ODX scores (i.e., 20-30) are more challenging to classify, likely due to the fact that 25 serves as the cut-off between low and high risk [5, 6]. On the other hand, this also indicates that the proposed model captures intrinsic features distinguishing ODX recurrence risk from the WSIs, although these features become ambiguous near the boundary between low and high risk. Our finding could also point to inherent ambiguity in these intermediate ODX scores from a clinical perspective.

## 4. Discussion and Conclusions

One of the most crucial steps in breast cancer prognosis is recurrence risk stratification. Accurate stratification enables healthcare providers to devise tailored treatment plans and optimize the use of chemotherapy, which will balance the benefits of treatment against potential risks and side effects. Receptor status plays a significant role in informing patients about their cancer progression, recurrence risk, and response to various treatments. Notably, HR+/HER2- cases constitute approximately 70% of breast cancer patients in the U.S. [34]. Patients with operable, HR+/HER2- breast cancer can often be treated effectively with endocrine therapies that are aimed at disrupting estrogen signaling, such as selective estrogen receptor modulators (tamoxifen) [35], or aromatase inhibitors (letrozole, anastrozole, or exemestane) [36, 37] if the recurrence risk is low. Conversely, patients with a high recurrence risk due to more aggressive disease biology, derive significant benefit from chemotherapy followed by endocrine therapy [38]. ODX examination is a gene assay commonly used to risk-stratify HR+/HER2- breast cancer patients. It has also been proven to be informative in assessing the efficacy (or lack of it) of chemotherapy [5]. However, due to the high cost of ODX, there is a notable disparity in its utilization among different racial/ethnic groups and around the world. This disparity underscores the need for more accessible and cost-effective prognostic tools.

Compared to gene expression assays, information gained from histopathology assessments is much more cost-effective and incorporated as a standard-of-care assessment tool for cancer diagnosis. In recent years, several computational pathology models have been proposed to predict ODX outcomes based on WSIs [18, 19, 39]. However, most of these methods rely on hand-crafted features and involve multiple steps, which can introduce human bias into their models. Recently, weakly supervised learning approaches have emerged as the mainstream in computational pathology for WSI analysis [23, 40-42]. Unlike traditional machine learning methods, weakly supervised learning models do not require extensive tissue annotations on WSIs but only a single diagnostic label for each WSI. This characteristic allows computational pathology models to overcome data limitations, enabling the training of models on a larger scale. Despite this progress, the application of weakly supervised learning models to predict ODX from WSIs has been limited, with existing models achieving only moderate performance [14]. In this study, we propose Deep-BCR-Auto as a fully automatic ODX prediction pipeline for H&E-stained WSIs. By leveraging the advantages of weakly supervised learning, Deep-BCR-Auto aims to enhance the accuracy and scalability of ODX predictions, providing a more efficient and unbiased approach to breast cancer prognosis.

We first trained and tested Deep-BCR-Auto on the publicly available TCGA-BRCA dataset. Then, we performed independent testing on an in-house dataset collected from The Ohio State University Wexner Medical Center (OSUWMC). Our three-fold cross-validation results on the TCGA-BRCA dataset are mainly reported in Table 2 and Figure 2. We found that the Deep-BCR-Auto achieved the best performance (AUROC 0.827, 95% CI 0.821-0.831) that outperforms CLAM-CTrans (AUROC 0.794, 95% CI 0.789-0.803) and CLAM (AUROC 0.750, 95% CI 0.742-0.756). By adjusting the threshold, we also found that our model achieved reasonable specificities while maintaining high sensitivities. Additionally, our model demonstrated excellent negative predictive values (NPVs). A high sensitivity model is critical in clinical practice to identify the majority of high-risk patients. Our model has exhibited its ability to precisely rule out low-risk patients, thereby reducing unnecessary ODX examinations. This property will make Deep-BCR-Auto a valuable tool in clinical settings, improving the efficiency of breast cancer prognosis and potentially lowering healthcare costs by minimizing the reliance on expensive and labor-intensive gene expression assays.

In our independent testing experiment, we selected the best-performing Deep-BCR-Auto model from the TCGA-BRCA experiments and evaluated it on the OSU dataset. The primary results are presented in Figure 5 and Table 4. The Deep-BCR-Auto model achieved an AUROC of 0.832 (95% CI 0.829-0.833), comparable to its performance on the TCGA-BRCA dataset. Additionally, after tuning the classification threshold using a small validation set from the OSU dataset, the model demonstrated strong classification performance on the OSU testing set (accuracy 82.0%, specificity 85.0%, sensitivity 67.7%). These findings further highlight the robust generalizability of the proposed model across data from different hospitals. The model's calibration curve in Figure 5c, particularly the slope of 0.62, indicates a slight underestimation of risk for high-risk patients, which aligns with the model's sensitivity. This reveals a potential area for future refinement, as improving the accuracy for high-risk patients could significantly impact clinical decision-making. From the other hand, the model demonstrates strong discrimination between high- and low-risk patients, evidenced by a C (ROC) of 0.83. Overall, the results demonstrate that Deep-BCR-Auto effectively predicts the ODX recurrence risk from the WSIs.

In our analysis of the Deep-BCR-Auto's performance on the OSU testing set, we observed that misclassifications predominantly occurred in cases with ODX scores in the 20 to 30 range (Figure 6). In contrast, samples with very low or high ODX scores were less frequently misclassified. These results highlight the challenge of accurately classifying cases with intermediate ODX scores, particularly around the threshold of 25, which delineates low and high-risk groups. This result also demonstrates that the proposed model effectively captures the intrinsic features related to ODX recurrence risk. However, the features are ambiguous when the ODX score is closer to the cut-off value.

Moreover, our study shows that a computational pathology model can achieve comparable prediction performances among African American, Caucasian, and Asian patients. This finding is particularly significant given that African American patients have been reported to have lower chances of receiving ODX due to cost barriers and healthcare disparities [9]. The ability of Deep-BCR-Auto to provide accurate and equitable prognostic information across different racial groups highlights the model's robustness and generalizability. By addressing these disparities, our results underscore the importance of using advanced computational methods to ensure all patients have access to high-quality breast cancer care. This uniformity in diagnostic capability ensures that all demographic groups receive the same high standard of care, potentially reducing racial and ethnic disparities in breast cancer outcomes. Consequently, Deep-BCR-Auto represents a major advancement in personalized medicine, potentially enhancing treatment planning and improving prognostic accuracy for diverse patient populations.

Although our study shows promising results, there are areas that could be improved. Our datasets exhibit a categorical imbalance between low and high-risk cases, particularly for the HR+/HER2- subset we focused on. As a result, our testing data includes a limited number of high-risk slides (~15% of all slides), which restricts our ability to evaluate the performance of our model thoroughly. This data imbalance could also impact the training process, causing the deep learning model to be biased toward predicting low-risk cases more frequently. This is expected, as high-risk cases are typically rare in ODX examinations. According to the National Cancer Database, only 15% of patients are classified as high-risk, which aligns with our data distribution [43]. Despite this, our proposed model achieved a balanced specificity, sensitivity, and excellent AUROC. In future studies, we plan to collect a larger cohort of patients to strengthen our dataset and further validate our model's performance. Additionally, we envision developing Deep-BCR-Auto into a web-based platform that can analyze ODX scores from uploaded WSIs, thereby extending access to this critically important test to regions where the actual test is currently unavailable.

In conclusion, Deep-BCR-Auto exhibits the potential to enhance personalized treatment planning and improve patient outcomes in breast cancer care. Future research should focus on validating our findings with larger clinical-level datasets and exploring the integration of this technology into routine clinical practice. By doing so, we can ensure that more patients benefit from precise and equitable breast cancer prognosis and treatment.

**Declaration of Competing Interest**

The authors declare no competing interests.

**Acknowledgments**

The project described was supported in part by R21 CA273665 (PIs: Gurcan) from the National Cancer Institute, R01 CA276301 (PIs: Niazi, Chen) from the National Cancer Institute, R21 EB029493 (PIs: Niazi, Segal) from the National Institute of Biomedical Imaging and Bioengineering, and GR125886 (PIs: Frankel, Niazi) from the Alliance Clinical Trials in Oncology. The content is solely the responsibility of the authors and does not necessarily represent the official views of the National Institutes of Health, National Institute of Biomedical Imaging and Bioengineering, Alliance Clinical Trials in Oncology, and National Cancer Institute.

**Data sharing statement**

Our codes and recurrence risk labels of TCGA-BRCA dataset are available at https://github.com/cialab/Deep-BCR-Auto. The TCGA-BRCA dataset is available at https://portal.gdc.cancer.gov/.

**Supplemental Methods**

1. Training method of patch classifier for tumor bulk segmentation

The developed patch classifier is a Swin Transformer model [44] initialized with pre-trained weights from large scale self-supervised training on histopathology images [29]. We appended a dropout layer and a randomly initialized linear classification layer after the global pooling layer of Swin Transformer. For this tumor/normal classification problem, we utilized binary cross-entropy loss function and Adam optimizer during fine-tuning on our own dataset. Our learning rate was 0.0001 and the weight decay parameter for Adam was 0.00005. During fine-tuning, the last transformer block and the linear classification layer are tunable. We set the maximum training duration as 100 epochs and early-stopped the training if the validation loss didn't decrease for five consecutive epochs.

Using 50 annotated WSIs, we randomly cropped 36,077 tumor patches and 36,077 normal patches, each sized 896×896 at 40× magnification. We conducted five-fold cross-validation with a hold-out testing set, using an 80/5/15 split at the WSI level (i.e., patches from each split are sourced from different WSIs). The model with the highest accuracy (89.2%) on the hold-out testing set was selected as the final patch classifier model.

## 2. Training method of Deep-BCR model for ODX recurrence risk prediction

To train the Deep-BCR model, we utilized binary cross-entropy loss function and Adam optimizer. Our learning rate was 0.0001 and the weight decay parameter for Adam was 0.00005. During the training, we set the maximum duration as 50 epochs and early stopped the training if the validation loss didn't decrease for five consecutive epochs. Additionally, while loading data for training, we used category-wise weighted sampling, assigning a higher sampling rate to high-risk samples compared to low-risk samples. This approach helped balance the number of high- and low-risk samples seen by the model in each epoch.

**References**

[1]  R. L. Siegel, A. N. Giaquinto, and A. Jemal, "Cancer statistics, 2024," *CA: A Cancer Journal for Clinicians,* 2024.
[2]  B. N. Joe. (2019, 8 February 2022). *Clinical features, diagnosis, and staging of newly diagnosed breast cancer.* Available: https://www.uptodate.com/contents/clinical-features-diagnosis-and-staging-of-newly-diagnosed-breast-cancer
[3]  N. Howlader, S. F. Altekruse, C. I. Li, V. W. Chen, C. A. Clarke, L. A. Ries, and K. A. Cronin, "US incidence of breast cancer subtypes defined by joint hormone receptor and HER2 status," *Journal of the National Cancer Institute,* vol. 106, no. 5, p. dju055, 2014.
[4]  S. Paik, S. Shak, G. Tang, C. Kim, J. Baker, M. Cronin, F. L. Baehner, M. G. Walker, D. Watson, T. Park, W. Hiller, E. R. Fisher, D. L. Wickerham, J. Bryant, and N. Wolmark, "A multigene assay to predict recurrence of tamoxifen-treated, node-negative breast cancer," *N Engl J Med,* vol. 351, no. 27, pp. 2817-26, Dec 30 2004.
[5]  J. A. Sparano, R. J. Gray, D. F. Makower, K. I. Pritchard, K. S. Albain, D. F. Hayes, C. E. Geyer Jr, E. C. Dees, M. P. Goetz, and J. A. Olson Jr, "Adjuvant chemotherapy guided by a 21-gene expression assay in breast cancer," *New England Journal of Medicine,* vol. 379, no. 2, pp. 111-121, 2018.
[6]  K. Kalinsky, W. E. Barlow, J. R. Gralow, F. Meric-Bernstam, K. S. Albain, D. F. Hayes, N. U. Lin, E. A. Perez, L. J. Goldstein, S. K. L. Chia, S. Dhesy-Thind, P. Rastogi, E. Alba, S. Delaloge, M. Martin, C. M. Kelly, M. Ruiz-Borrego, M. Gil-Gil, C. H. Arce-Salinas, E. G. C. Brain, E. S. Lee, J. Y. Pierga, B. Bermejo, M. Ramos-Vazquez, K. H. Jung, J. M. Ferrero, A. F. Schott, S. Shak, P. Sharma, D. L. Lew, J. Miao, D. Tripathy, L. Pusztai, and G. N. Hortobagyi, "21-Gene Assay to Inform Chemotherapy Benefit in Node-Positive Breast Cancer," (in eng), *N Engl J Med,* vol. 385, no. 25, pp. 2336-2347, Dec 16 2021.
[7]  T. w. bank. (2023). Available: https://datatopics.worldbank.org/world-development-indicators/the-world-by-income-and-region.html
[8]  J. Moore, F. Wang, T. Pal, S. Reid, H. Cai, C. E. Bailey, W. Zheng, L. Lipworth, and X.-O. Shu, "Oncotype DX risk recurrence score and total mortality for early-stage breast cancer by race/ethnicity," *Cancer Epidemiology, Biomarkers & Prevention,* vol. 31, no. 4, pp. 821-830, 2022.
[9]  M. C. Roberts, M. Weinberger, S. B. Dusetzina, M. A. Dinan, K. E. Reeder-Hayes, L. A. Carey, M. A. Troester, and S. B. Wheeler, "Racial variation in the uptake of oncotype DX testing for early-stage breast cancer," *Journal of Clinical Oncology,* vol. 34, no. 2, p. 130, 2016.
[10] L. J. Collin, M. Yan, R. Jiang, K. C. Ward, B. Crawford, M. A. Torres, K. Gogineni, P. D. Subhedar, S. Puvanesarajah, and M. M. Gaudet, "Oncotype DX recurrence score implications for disparities in chemotherapy and breast cancer mortality in Georgia," *NPJ breast cancer,* vol. 5, no. 1, p. 32, 2019.

[11]  L. J. Ricks-Santi and J. T. McDonald, "Low utility of Oncotype DX® in the clinic," *Cancer Medicine,* vol. 6, no. 3, pp. 501-507, 2017.
[12]  A. Dhungana, A. Vannier, F. Zhao, J. Q. Freeman, P. Saha, M. Sullivan, K. Yao, E. M. Flores, O. I. Olopade, and A. T. Pearson, "Development and validation of a clinical breast cancer tool for accurate prediction of recurrence," *npj Breast Cancer,* vol. 10, no. 1, p. 46, 2024.
[13]  Z. Su, A. Rosen, R. Wesolowski, G. Tozbikian, M. K. K. Niazi, and M. N. Gurcan, "Deep-ODX: an efficient deep learning tool to risk stratify breast cancer patients from histopathology images," in *Medical Imaging 2024: Digital and Computational Pathology*, 2024, vol. 12933, pp. 34-39: SPIE.
[14]  Z. Su, M. K. K. Niazi, T. E. Tavolara, S. Niu, G. H. Tozbikian, R. Wesolowski, and M. N. Gurcan, "BCR-Net: A deep learning framework to predict breast cancer recurrence from histopathology images," (in eng), *PLoS One,* vol. 18, no. 4, p. e0283562, 2023.
[15]  H. Li, Y. Zhu, E. S. Burnside, K. Drukker, K. A. Hoadley, C. Fan, S. D. Conzen, G. J. Whitman, E. J. Sutton, J. M. Net, M. Ganott, E. Huang, E. A. Morris, C. M. Perou, Y. Ji, and M. L. Giger, "MR Imaging Radiomics Signatures for Predicting the Risk of Breast Cancer Recurrence as Given by Research Versions of MammaPrint, Oncotype DX, and PAM50 Gene Assays," *Radiology,* vol. 281, no. 2, pp. 382-391, 2016.
[16]  F. M. Howard, J. Dolezal, S. Kochanny, G. Khramtsova, J. Vickery, A. Srisuwananukorn, A. Woodard, N. Chen, R. Nanda, C. M. Perou, O. I. Olopade, D. Huo, and A. T. Pearson, "Integration of clinical features and deep learning on pathology for the prediction of breast cancer recurrence assays and risk of recurrence," *npj Breast Cancer,* vol. 9, no. 1, p. 25, 2023/04/14 2023.
[17]  T. E. Tavolara, Z. Su, M. N. Gurcan, and M. K. K. Niazi, "One label is all you need: Interpretable AI-enhanced histopathology for oncology," in *Seminars in Cancer Biology*, 2023: Elsevier.
[18]  J. Whitney, G. Corredor, A. Janowczyk, S. Ganesan, S. Doyle, J. Tomaszewski, M. Feldman, H. Gilmore, and A. Madabhushi, "Quantitative nuclear histomorphometry predicts oncotype DX risk categories for early stage ER+ breast cancer," *BMC cancer,* vol. 18, no. 1, pp. 1-15, 2018.
[19]  H. Li, J. Whitney, K. Bera, H. Gilmore, M. A. Thorat, S. Badve, and A. Madabhushi, "Quantitative nuclear histomorphometric features are predictive of Oncotype DX risk categories in ductal carcinoma in situ: preliminary findings," *Breast cancer research,* vol. 21, no. 1, pp. 1-16, 2019.
[20]  J. Yang, J. Ju, L. Guo, B. Ji, S. Shi, Z. Yang, S. Gao, X. Yuan, G. Tian, and Y. Liang, "Prediction of HER2-positive breast cancer recurrence and metastasis risk from histopathological images and clinical information via multimodal deep learning," *Computational and structural biotechnology journal,* vol. 20, pp. 333-342, 2022.
[21]  Y. Chen, H. Li, A. Janowczyk, P. Toro, G. Corredor, J. Whitney, C. Lu, C. F. Koyuncu, M. Mokhtari, C. Buzzy, S. Ganesan, M. D. Feldman, P. Fu, H. Corbin, A. Harbhajanka, H. Gilmore, L. J. Goldstein, N. E. Davidson, S. Desai, V. Parmar, and A. Madabhushi, "Computational pathology improves risk stratification of a multi-gene assay for early stage ER+ breast cancer," *npj Breast Cancer,* vol. 9, no. 1, p. 40, 2023/05/17 2023.
[22]  Z. Su, M. Rezapour, U. Sajjad, M. N. Gurcan, and M. K. K. Niazi, "Attention2Minority: A salient instance inference-based multiple instance learning for classifying small lesions in whole slide images," *Computers in Biology and Medicine,* vol. 167, p. 107607, 2023/12/01/ 2023.
[23]  H. Zhang, Y. Meng, Y. Zhao, Y. Qiao, X. Yang, S. E. Coupland, and Y. Zheng, "DTFD-MIL: Double-Tier Feature Distillation Multiple Instance Learning for Histopathology Whole Slide Image Classification," in *Proceedings of the IEEE/CVF Conference on Computer Vision and Pattern Recognition*, 2022, pp. 18802-18812.
[24]  W. Tang, S. Huang, X. Zhang, F. Zhou, Y. Zhang, and B. Liu, "Multiple instance learning framework with masked hard instance mining for whole slide image classification," in *Proceedings of the IEEE/CVF International Conference on Computer Vision*, 2023, pp. 4078-4087.
[25]  R. J. Chen, M. Y. Lu, D. F. Williamson, T. Y. Chen, J. Lipkova, Z. Noor, M. Shaban, M. Shady, M. Williams, and B. Joo, "Pan-cancer integrative histology-genomic analysis via multimodal deep learning," *Cancer Cell,* vol. 40, no. 8, pp. 865-878. e6, 2022.
[26]  A. Orucevic, J. L. Bell, M. King, A. P. McNabb, and R. E. Heidel, "Nomogram update based on TAILORx clinical trial results-Oncotype DX breast cancer recurrence score can be predicted using clinicopathologic data," *The Breast,* vol. 46, pp. 116-125, 2019.
[27]  M. Y. Lu, D. F. Williamson, T. Y. Chen, R. J. Chen, M. Barbieri, and F. Mahmood, "Data-efficient and weakly supervised computational pathology on whole-slide images," *Nature biomedical engineering,* vol. 5, no. 6, pp. 555-570, 2021.

[28] G. Campanella, M. G. Hanna, L. Geneslaw, A. Miraflor, V. W. K. Silva, K. J. Busam, E. Brogi, V. E. Reuter, D. S. Klimstra, and T. J. Fuchs, "Clinical-grade computational pathology using weakly supervised deep learning on whole slide images," *Nature Medicine,* vol. 25, no. 8, pp. 1301-+, Aug 2019.

[29] X. Wang, S. Yang, J. Zhang, M. Wang, J. Zhang, W. Yang, J. Huang, and X. Han, "Transformer-based unsupervised contrastive learning for histopathological image classification," *Medical Image Analysis,* vol. 81, p. 102559, 2022.

[30] E. Reinhard, M. Adhikhmin, B. Gooch, and P. Shirley, "Color transfer between images," *IEEE Computer Graphics and Applications,* vol. 21, no. 5, pp. 34-41, 2001.

[31] M. Kim, H. Sekiya, G. Yao, N. B. Martin, M. Castanedes-Casey, D. W. Dickson, T. H. Hwang, and S. Koga, "Diagnosis of Alzheimer disease and tauopathies on whole-slide histopathology images using a weakly supervised deep learning algorithm," *Laboratory investigation,* vol. 103, no. 6, p. 100127, 2023.

[32] Z. Chen, I. H. Wong, W. Dai, C. T. Lo, and T. T. Wong, "Lung Cancer Diagnosis on Virtual Histologically Stained Tissue Using Weakly Supervised Learning," *Modern Pathology,* vol. 37, no. 6, p. 100487, 2024.

[33] Q. Zeng, C. Klein, S. Caruso, P. Maille, N. G. Laleh, D. Sommacale, A. Laurent, G. Amaddeo, D. Gentien, and A. Rapinat, "Artificial intelligence predicts immune and inflammatory gene signatures directly from hepatocellular carcinoma histology," *Journal of Hepatology,* vol. 77, no. 1, pp. 116-127, 2022.

[34] N. C. Institute. *Bethesda MD SEER Cancer Stat Facts: Female Breast Cancer*. Available: https://seer.cancer.gov/statfacts/html/breast.html

[35] M. Untch and C. Thomssen, "Clinical practice decisions in endocrine therapy," *Cancer Invest,* vol. 28 Suppl 1, pp. 4-13, 2010.

[36] M. M. Regan, P. Neven, A. Giobbie-Hurder, A. Goldhirsch, B. Ejlertsen, L. Mauriac, J. F. Forbes, I. Smith, I. Láng, A. Wardley, M. Rabaglio, K. N. Price, R. D. Gelber, A. S. Coates, and B. Thürlimann, "Assessment of letrozole and tamoxifen alone and in sequence for postmenopausal women with steroid hormone receptor-positive breast cancer: the BIG 1-98 randomised clinical trial at 8·1 years median follow-up," *The Lancet Oncology,* vol. 12, no. 12, pp. 1101-1108, 2011/11/01/ 2011.

[37] H. J. Burstein and J. J. Griggs, "Adjuvant hormonal therapy for early-stage breast cancer," *Surg Oncol Clin N Am,* vol. 19, no. 3, pp. 639-47, Jul 2010.

[38] J. Anampa, D. Makower, and J. A. Sparano, "Progress in adjuvant chemotherapy for breast cancer: an overview," *BMC Med,* vol. 13, p. 195, Aug 17 2015.

[39] D. Romo-Bucheli, A. Janowczyk, H. Gilmore, E. Romero, and A. Madabhushi, "Automated tubule nuclei quantification and correlation with oncotype DX risk categories in ER+ breast cancer whole slide images," *Scientific reports,* vol. 6, no. 1, pp. 1-9, 2016.

[40] M. Y. Lu, D. F. K. Williamson, T. Y. Chen, R. J. Chen, M. Barbieri, and F. Mahmood, "Data-efficient and weakly supervised computational pathology on whole-slide images," *Nature Biomedical Engineering,* vol. 5, no. 6, pp. 555-570, 2021/06/01 2021.

[41] R. J. Chen, C. Chen, Y. Li, T. Y. Chen, A. D. Trister, R. G. Krishnan, and F. Mahmood, "Scaling vision transformers to gigapixel images via hierarchical self-supervised learning," in *Proceedings of the IEEE/CVF Conference on Computer Vision and Pattern Recognition*, 2022, pp. 16144-16155.

[42] L. Fillioux, J. Boyd, M. Vakalopoulou, P.-H. Cournède, and S. Christodoulidis, "Structured state space models for multiple instance learning in digital pathology," in *International Conference on Medical Image Computing and Computer-Assisted Intervention*, 2023, pp. 594-604: Springer.

[43] K. Kalinsky, W. E. Barlow, J. R. Gralow, F. Meric-Bernstam, K. S. Albain, D. F. Hayes, N. U. Lin, E. A. Perez, L. J. Goldstein, and S. K. Chia, "21-gene assay to inform chemotherapy benefit in node-positive breast cancer," *New England Journal of Medicine,* vol. 385, no. 25, pp. 2336-2347, 2021.

[44] Z. Liu, Y. Lin, Y. Cao, H. Hu, Y. Wei, Z. Zhang, S. Lin, and B. Guo, "Swin transformer: Hierarchical vision transformer using shifted windows," in *Proceedings of the IEEE/CVF international conference on computer vision*, 2021, pp. 10012-10022.